\input{aipcheck}

\documentclass[
    ,final            
  ]
  {aipproc}

\layoutstyle{6x9}

\begin{document}
\newcommand{\begit}{\begin{itemize}}
\newcommand{\enit}{\end{itemize}}
\newcommand{\begen}{\begin{enumerate}}
\newcommand{\enen}{\end{enumerate}}
\newcommand{\beq}{\begin{equation}} 
\newcommand{\eeq}{\end{equation}} 
\newcommand{\beqa}{\begin{eqnarray}} 
\newcommand{\eeqa}{\end{eqnarray}} 
\def\lesssim{\mathrel{\hbox{\rlap{\hbox{\lower4pt\hbox{$\sim$}}}\hbox{$<$}}}}
\def\gtrsim{\mathrel{\hbox{\rlap{\hbox{\lower4pt\hbox{$\sim$}}}\hbox{$>$}}}}

\title{Proto-Magnetars as GRB Central Engines: 
Uncertainties, Limitations, \& Particulars}


\author{Todd A. Thompson}{
  address={Department of Astronomy and Center
for Cosmology \& Astro-Particle Physics, 
The Ohio State University, Columbus, Ohio 43210; 
thompson@astronomy.ohio-state.edu},
altaddress={Alfred P.~Sloan Fellow}
}

\author{Brian D.~Metzger}{
address=
{Department of Astrophysical Sciences, Peyton Hall, Princeton Univ.; Princeton, NJ 08544 USA},
altaddress={Einstein Fellow}}

\author{Niccol\`o Bucciantini}{
address={NORDITA, Roslagstullsbacken 23, 106 91 Stockholm, Sweden}}

\begin{abstract}
The millisecond proto-magnetar model for the central engine 
of long-duration gamma-ray bursts is briefly reviewed.  Limitations and
uncertainties in the model are highlighted.  A short discussion of the 
maximum energy, maximum duration, radiative efficiency,
jet formation mechanism, late-time energy injection, and (non-)association with supernovae
of millisecond magnetar-powered GRBs is provided.
\end{abstract}

\maketitle


\section{Introduction}

Models for the central engine of long-duration gamma ray 
bursts (GRBs) are highly constrained by the character of  
the prompt emission and the afterglow, 
and --- at least in some cases (and perhaps nearly all \cite{woosley_bloom})
--- the fact of an associated supernova (SN).  
A successful model must at minimum satisfy several criteria. It must generate 
a collimated outflow with high Lorentz factor 
$\gamma_\infty\gtrsim10^2-10^3$
and high kinetic luminosity ($\dot{E}\sim10^{50}-10^{51}$\,ergs s$^{-1}$) 
on a timescale of $\sim10-100$ seconds.  Additionally,
inferences from X-ray observations by {\it Swift} indicate that there
may be late-time energy injection by the central engine on timescales
of hours to days \cite{dburrows}.  

There is little diversity among models for the central 
engine and essentially all can be simply classed as a rotating 
compact object that drives an asymmetric relativistic outflow. 
The two leading models are the ``collapsar'' and the ``millisecond proto-magnetar.''
The former, as described in \cite{Woosley93} and developed
in \cite{MW99,macfadyen_woosley_heger}, proposes that GRBs arise from the collapse of rapidly 
rotating Type-Ibc progenitors. A black hole forms with an accompanying
accretion disk and drives a collimated relativistic outflow along the 
axis of rotation via either neutrino heating or magnetic stresses.
Ref \cite{dessart_collapsar} have recently pointed out that rapid mass
loss during rotating collapse might prevent the formation
of a central black hole, at least for some time.

In contrast to the collapsar mechanism, 
the ``millisecond proto-magnetar'' model posits a newly formed 
rapidly rotating neutron star  (spin period $P\sim1$\,ms) 
with surface magnetic field of magnetar strength 
($B\gtrsim10^{15}$\,G), cooling via neutrino radiation, and driving
a neutrino-heated magneto-centrifugal wind \cite{thompson04,bucciantini06,M07a}.
Millisecond proto-magnetars might be produced by rotating Type-II and 
Type-Ibc progenitors, the accretion-induced collapse of a white dwarf,
the merger of two white dwarfs, and/or (potentially) the 
merger of two neutron stars \cite{MQT07}. Thus, 
they may trace both young and old stellar populations.
See  Refs \cite{usov,thompson94,wheeler,uzdensky_macfadyen} for more on the
 development of this model.

Here, we briefly review the millisecond proto-magnetar
model for long-duration GRBs.  We then discuss some
of the particulars of the model, including some
of its limitations and uncertainties.
A recent comparison between the collapsar and millisecond-magnetar
models can be found in Ref \cite{metzger_review}.

\vspace{-0.2cm}
\section{The Birth of Millisecond Proto-Magnetars}
\vspace{-0.2cm}

Core-collapse supernovae (SNe) leave behind hot 
proto-neutron stars that cool on the Kelvin-Helmholtz 
timescale ($t_{\rm KH}\approx10-100$ s) by
radiating their gravitational binding energy ($\sim10^{53.5}$ ergs) in neutrinos
\cite{burrows_lattimer,pons}.  
A fraction of these neutrinos deposit their energy in the tenuous 
and extended atmosphere of the proto-neutron star.\footnote{The charged-current
interactions $\nu_e n\leftrightarrow p e^-$ and $\bar{\nu}_e p\leftrightarrow n e^+$
dominate heating and cooling.}  In the standard picture, net neutrino heating drives a thermal 
wind that emerges into the post-SN shock environment,
blowing a wind-driven bubble into the exploding and expanding SN cavity 
\cite{woosley94,bhf}.  
A neutrino-driven wind is
obtained in all successful models of SNe, regardless of
how the explosion is seeded and is a natural consequence of the 
low-pressure cavity created by
the explosion and the very high thermal pressure at the 
proto-neutron star surface
\cite{bhf,janka_muller,scheck}.\footnote{This statement is true even if the 
wind is ``one-sided'' in its first seconds, 
as in Ref \cite{scheck}.}
For most massive stellar
progenitors with extended hydrogen envelopes (Type-II), the cooling phase is over well
before shock breakout at the progenitor's surface ($\sim$\,1 hour after collapse).  
For compact Type-Ibc SNe, the SN shockwave traverses the progenitor 
on a timescale  comparable to $t_{\rm KH}$.

 For a neutron star with $M=1.4$\,M$_\odot$,
$R=10$\,km, rotation frequency $\Omega$, and 
a magnetar-strength magnetic field, the mass-loss rate 
during $t_{\rm KH}$ is given by \cite{duncan_shapiro_wasserman,QW96,thompson04,M07a}
\beq
\dot{M}(t)\approx4\times10^{-7}L^{5/2}_{\nu,\,51.5}(t)
\exp[\Omega(t)^2/\Omega_{\rm c}^2]\,\,\,{\rm M_\odot \,\,\,s^{-1}},
\label{mdot}
\eeq
where $L_{\nu, \,51.5}(t)=L_\nu(t)/10^{51.5}$ ergs s$^{-1}$ 
is the total neutrino luminosity,
and $L_\nu\propto\langle \varepsilon_\nu\rangle^4$ has been assumed,
where $\langle \varepsilon_\nu\rangle$ is the average
neutrino energy.  The normalization of 
equation (\ref{mdot}) and its $L_\nu$ dependence follow
from the physics of the weak interaction and the depth
of the NS gravitational potential
\cite{duncan_shapiro_wasserman,QW96}.  
The exponential factor in equation (\ref{mdot})
accounts for the mass loss enhancement by magneto-centrifugal 
forces when $B$ and $\Omega$ are large 
($\Omega_c\approx2000L_{\nu,\,51.5}^{0.08}$\,rad s$^{-1}$
\cite{thompson04,M07a}).

One of the most important components of the proto-magnetar model is that as the 
neutron star cools, $L_\nu(t)$ and $\langle\varepsilon_\nu\rangle(t)$
decrease on a timescale $t_{\rm KH}$. 
Typically, the dependence $L_\nu(t)\propto t^{-1}$ is found in
cooling models of non-rotating non-magnetic proto-neutron stars 
until $t\sim30-40$\,s when $L_\nu(t)$ plummets as the neutron
star becomes transparent \cite{pons}.
As a result of equation 
(\ref{mdot}), as $L_\nu(t)$ decreases, $\dot{M}(t)$ decreases. 
For this reason, for fixed surface magnetic field strength $B$, 
the wind becomes increasingly magnetically-dominated and relativistic as a function of time.
Like beads on a stiff wire, the matter is accelerated off of the
surface of the proto-neutron star by the combined action of 
big $B$ and $\Omega$. 
The degree to which the magnetic field dominates the dynamics
and accelerates the flow
is quantified by the magnetization at the light cylinder ($R_L=c/\Omega\sim50P_1$\,km):
\beq
\sigma_{\rm LC} = \left.B^2/(4\pi\rho c^2)\right|_{\rm LC}
\approx75 \,B_{15}^2 P_1^{-4}L^{-5/2}_{\nu,\,51.5}(t),
\label{sigma}
\eeq
where $B_{15}=B/10^{15}$\,G is the surface dipole field strength.
The Lorentz factor of an outflow at large
distances can approach $\gamma_\infty \rightarrow \sigma_{\rm LC}$ (see Ref 
\cite{drenkhahn_spruit}) if there is
efficient conversion of magnetic energy into kinetic energy (in which
case $\sigma(r)$ itself will become small at large $r$).
Thus, under the assumption of efficient conversion of 
magnetic energy to kinetic energy $\gamma_\infty(t)$ increases
dramatically as $\dot{M}$ drops on $t_{\rm KH}$.

Starting from a time $t\sim1$\,s
after the collapse and successful SN explosion, there are
several phases in the life a proto-neutron star wind.  Although the 
specific timing of each phase depends on $B$ and $\Omega$, for a 
millisecond magnetar they are roughly the following.  For a second
or so after explosion, as the proto-neutron star contracts to its
final radius and spins up, the wind is pressure-dominated and 
driven by neutrino energy deposition.  The asymptotic velocity
of the flow is $\sim0.1c$.  Just a few seconds
later, the wind rapidly becomes magneto-centrifugally
dominated, but it is still non-relativistic ($\sigma_{\rm LC}<1$).  
Perhaps $\sim2-5$\,s into the cooling epoch,  
$\dot{M}(t)$ decreases sufficiently that $\sigma_{\rm LC}(t)$
becomes larger than unity, the flow is accelerated to near $c$
by magneto-centrifugal forces from the surface of the neutron
star out to the light cylinder.
It is this late-time outflow --- deep into the cooling epoch of the magnetar,  
many seconds after the explosion is initiated --- that is the most 
promising for producing a GRB.  Eventually, at the end of 
the cooling epoch $L_\nu(t)$ and $\dot{M}(t)$ decrease dramatically
(see Ref \cite{pons}), 
$\sigma(t)$ increases to $\sim10^6$ and the magneto-hydrodynamical 
mass loss ceases, and the millisecond magnetar enters its
``pulsar'' phase in which its energy loss rate is presumed
to be given by the force-free (``vacuum dipole'') spindown formula 
\cite{spitkovsky}
\beq \dot{E}_{\rm
FF}\approx B^2 R^6 \Omega^4/c^3.
\label{edot_ff}
\eeq

The wind does not emerge into vacuum.  It follows the preceding,
SN shockwave, which propagates through
the progenitor at $\sim10,000$\,km s$^{-1}$.  As the wind becomes
increasingly fast a few seconds after the explosion is initiated, it shocks on the
slower outgoing SN shockwave.  This interaction 
produces a wind-driven ``magnetar wind nebula,'' and the residual azimuthal 
component of the magnetic field within the cavity exerts a ``pinch'' force 
that leads to a prominent and strong jet that punches 
through the star \cite{bucciantini07,bucciantini08,bucciantini09}.  
This effect is equivalent to that
discussed in Refs \cite{begelman_li, konigl_granot}.
See also \cite{uzdensky_macfadyen}.  

\section{Uncertainties, Limitations, \& Some Particulars}

\vspace*{-.05cm}
\subsection{Efficiency \& Internal Shocks}
\vspace*{-.2cm}

As discussed above, $\sigma_{\rm LC}(t)$ and (potentially)
$\gamma_\infty(t)$ increases monotonically as a function of
time in the proto-magnetar model.  Although there may be 
eruptions from the magnetar surface during cooling which
rapidly modulate $\dot{M}$ and $\sigma$, the overall trend
from low $\sigma$ ($<1$) to high $\sigma$ ($>10^5$) on a
$\sim10-100$\,s timescale is unavoidable.  If $\sigma(t)\propto\gamma_\infty(t)$,
this leads to very high radiative efficiency via internal
shocks (approaching $\sim0.3$; see  Ref \cite{MQT07}
for discussion in the context of  060614-like GRBs \cite{galyam,gehrels}).

\vspace*{-.5cm}
\subsection{Jet Formation}
\vspace*{-.2cm}

Perhaps counter-intuitively, 
the models of Refs \cite{bucciantini07,bucciantini08,bucciantini09}
 show that the otherwise predominantly equatorial proto-magnetar wind can be 
efficiently converted into a jet.  Importantly, these works find that 
little of spindown (wind) power is transferred to the ``spherical'' SN
component of the explosion.  While the slow SN shock acts to 
contain the wind and allow the jet to develop, a variety of simulations
with different parameters ($B$ and $\Omega$) show that the 
majority of the wind power escapes along the axis of these
axisymmetric relativistic MHD simulations.  This property
of the system preserves the overall degree of relativity of the
outflow from the light cylinder to the edge of the progenitor
(no spindown energy is ``lost'' to the quasi-spherical SN).
It is then difficult to argue that 
the relativistic magnetar outflow acts to super-energize the
SN explosion for comparison with events like SN 2003dh (GRB 030329;
\cite{stanek03}), which may require Ni yields and ejecta velocities 
higher than typical Ic SNe \cite{bucciantini09}.

Several theoretical questions remain: (1) What happens to the 
process of jet formation in 3D, when the magnetic and spin axes
of the proto-magnetar may be misaligned?  (2) If jet formation is 
indeed robust, is the alternating 
field (``striped'' wind) configuration expected for an oblique rotator 
preserved along the pole in the jet?   That is, is the jet stable? 
This issue bears critically
on the efficacy of magnetic reconnection models for the jet 
acceleration and emission \cite{drenkhahn_spruit}.

\vspace*{-.5cm}
\subsection{The Maximum Energy}
\vspace*{-.2cm}

The gravitational binding energy of a neutron star is of
order $E_{\rm grav}\sim GM^2/R\sim5\times10^{53}$\,ergs.  For maximal
rotation, one might guess that the total amount of rotational
energy that can be stored in a proto-neutron star at birth is 
$\sim E_{\rm grav}$, and thus that the maximum possible 
energy for a GRB powered by the rotational energy of a neutron
star is $E_{\rm rot}\sim E_{\rm grav}$.  However, at very rapid rotation
rates we expect strong deformations of the neutron star
(as in, e.g.,  Ref \cite{ott_onearm}), which will emit
copious gravitational waves.  These losses will effectively
spin down the proto-magnetar.  For example, taking the kinetic
power in the outflow to be $\dot{E}_{\rm wind}\sim \dot{M} R^2\Omega^2/2$
as the flow starts to become magnetically-dominated, and taking
the energy loss rate in gravity waves to be $\dot{E}_{\rm GW}
=(32/5)GI^2\epsilon^2\Omega^6/c^5$ \cite{arons}, where $I=I_{45}10^{45}$\,cgs is the moment
of inertia and $\epsilon=\epsilon_{0.01}0.01$ is the ellipticity, the critical 
$\Omega$ above which $\dot{E}_{\rm GW}>\dot{E}_{\rm wind}$
is $\Omega\sim5000\dot{M}_{-3}^{1/4}R_{10\,{\rm km}}^{1/2}
I_{45}^{-1/2}\epsilon_{0.01}^{-1/2}$\,rad s$^{-1}$,
where $\dot{M}_{-3}=\dot{M}/10^{-3}$\,M$_\odot$ s$^{-1}$
and $R_{10\,{\rm km}}=R/10$\,km.  Note that the magnetic field
within the neutron star may produce non-zero $\epsilon$ \cite{cutler,stella}.
Although sophisticated models
of rapidly rotating neutron star birth are required to address 
this question in detail, this estimate and others
suggests that the magnetar can have a maximum total rotational 
energy  of $E_{\rm max}\sim5\times10^{52}$\,ergs at the time of cooling. 
Such a calculation is particularly relevant since recent 
work suggests that some GRBs may approach this bound
\cite{cenko2,cenko1}.  However, the inference of the total
energy of the GRB depends on observation of the jet ``break''
in the afterglow lightcurve.  Work by Ref \cite{vaneerten}
suggests earlier calculations of the total energetics 
via jet breaks may overestimate the energy in the GRB jet
by a factor of $\sim3-4$. 

A theoretical investigation of rapidly rotating 3D MHD
stellar collapse is required to address the issue of $E_{\rm max}$
for millisecond proto-magnetars in detail.  Although one might 
guess that there is no similar bound for the collapsar mechanism
(short of the entire rest mass energy of the accreted material),
Ref \cite{dessart_collapsar} have found that it might be difficult to 
promptly form very rapidly rotating black holes in stellar 
collapse because of rapid mass loss.  Nevertheless, $E_{\rm max}$
remains an important potential observational probe of the 
GRB central engine.  It is particularly interesting to speculate
that observations may one day provide evidence for a 
well-defined maximum energy for long-duration GRBs, thus
providing a crucial clue to their nature.

\vspace*{-.5cm}
\subsection{The Maximum Duration}
\vspace*{-.2cm}

Equation (\ref{sigma}) shows that $\sigma_{\rm LC}$ (and, hence, 
$\gamma_\infty$) is a strong function of 
$L_\nu$.  In turn, $L_\nu$ is a strong function of 
time in non-rotating models of proto-neutron star cooling 
\cite{burrows_lattimer,pons}.  This time evolution implies
that the wind quickly evolves from $\sigma\sim1$ to $\sigma\sim1000$
on a timescale of less than $\sim30$\,s.  If the prompt emission 
mechanism requires that $\sigma$ is less than some upper 
bound (say, 1000; see \cite{nakar}), then the sharp cutoff in $L_\nu$
at the end of the cooling phase should quickly end the
prompt emission.  This implies that proto-magnetar powered prompt
emission from GRBs should have a sharp temporal distribution in 
the rest frame (see \cite{butler}). 

In the magnetar model, can the duration of the
prompt emission be increased?  There are potentially two ways.
First, the neutrino diffusion timescale is longer for
more massive NSs.  For a 2\,M$_\odot$ NS, the time at which
$L_\nu$ suddenly decreases, increases somewhat, from 
$\sim30$\,s for a 1.4 to perhaps $40-50$\,s.  The second way
is through rapid rotation.  Although models of rapidly rotating
neutron star cooling have not yet been constructed, parameterized
1D models of core-collapse SNe with rotation 
indicate that the overall neutrino luminosities and 
average energies might be decreased by a factor of 
at most $\sim5-6$ \cite{tqb}.  Since, at zeroth order,
the same amount of gravitational binding energy must be 
liberated, we expect that $t_{\rm KH}$ should increase
by the same factor.  Thus, if an upper bound on $\sigma$
(or $\gamma_\infty$) determines when the prompt emission
shuts off, then the maximum duration of a long-duration
GRB powered by a ms-magnetar is no longer than $\sim200$\,s.
Observations of GRBs showing that the prompt phase lasts
more than this in the rest frame would strain the magnetar model.

\vspace*{-.5cm}
\subsection{Flares \& Late-Time Activity}
\vspace*{-.2cm}

At late times, after neutrino-driven mass loss from 
the proto-neutron star has subsided and deleptonization is 
complete, the energy loss rate of the magnetar should
approach the force-free limit, given in equation (\ref{edot_ff}).
This implies that $\dot{E}\propto\Omega^4\propto t^{-2}$.
However, we know from studies of pulsars in the Galaxy
that the observed braking index is $n\equiv\Omega\ddot\Omega/\dot\Omega^2\sim2.5-2.9$
(see \cite{bucciantini06}, and references therein).  While the cause
of $n\ne3$ braking index is unknown, this finding implies that we 
should expect expect $\dot{E}\propto t^{-2.33}$ to $\propto t^{-2.05}$ in the 
force-free pulsar-like phase, if newly born magnetars spin
down like their less-energetic pulsar cousins.  In any case, 
the rotational energy reservoir should likely decrease monotonically,
and this behavior should imprint itself on any 
emission during this phase.  
For this reason, the late-time flares have been
analyzed by Ref \cite{lazzati} and \cite{chincarini}.
The former find that the average flare luminosity declines as a
powerlaw in time as $L\propto t^{-1.5\pm0.16}$, whereas the 
latter find that the peak luminosity of the flares 
correlates with the timescale of the peak as 
$L\propto t^{-2.7\pm0.5}$.  The former is likely too
shallow a time-dependence for the magnetar model 
if the flares are interpreted as powered by spindown \cite{lazzati},
whereas the latter may be somewhat too steep.

Recent work by Ref \cite{dallosso} and \cite{yu_magnetar}
explicitly connect the X-ray plateaus seen in the 
early afterglow of many {\it Swift} long-duration GRBs
with magnetar spindown.  If this correspondence is correct,
the magnetar model should predict correlations between the 
character of the prompt emission and the duration and flux of 
the plateau.  

Also regarding very late-time central engine activity, it has 
been suggested that the discovery of a giant
magnetar flare (a scaled-up version of those that occur from 
Galactic magnetars) associated with an old GRB
afterglow in a nearby galaxy would provide a striking 
confirmation of the magnetar GRB model \cite{giannios}.

\vspace*{-.5cm}
\subsection{Time Evolution of the Magnetic Field}
\vspace*{-.2cm}

A complicating and uncertain feature of the magnetar
model for GRBs is the time evolution of the magnetic
field.  This could be important for both the early-time
and late-time evolution and emission.  At early times,
during the neutrino cooling phase, $B$ may evolve because
of dynamo action in the proto-magnetar during the explosion
itself, or during its convective cooling phase \cite{DT92,TD93}.
This effect could extend or shorten the prompt phase 
if a certain range of $\sigma_{\rm LC}$ (or $\gamma_\infty$) 
is required for efficient emission.  After the prompt phase, 
$B$ might evolve in the magnetar via magnetic dissipation \cite{pons_geppert},
which could potentially lead to extended late-time X-ray activity.
A time-variable $B$ would also  
complicate the time dependence of spindown at all phases
(eq.~\ref{edot_ff}).

\vspace*{-.5cm}
\subsection{(Non-)Association with Supernovae}
\vspace*{-.2cm}

Some proto-magnetars may be formed from the collapse of a
Type-Ibc progenitor.  In this scenario, one imagines 
that the SN mechanism in some form (e.g., the ``neutrino mechanism''
\cite{bethe_wilson}) launches a SN shock with $\sim10^{51}$\,ergs
that produces $^{56}$Ni via explosive nucleosynthesis.  This would
explain the presence of an accompanying SN, but not the bright
character of some GRB-SNe.  If  $^{56}$Ni 
in excess of that produced in non-GRB Type-Ibc SNe is indeed
required in some cases (see \cite{soderberg_SN}) 
there are two ways in which it might be 
produced by proto-magnetars: (1) some of 
the initial rotational energy of the core may be tapped rapidly
via magnetic stresses, enhancing the 
SN shock energy as it is launched \cite{tqb,burrowsB} 
and/or (2) as the initial slow SN shockwave is moving outward it is 
shocked by the subsequent, highly-energetic proto-magnetar 
wind, again enhancing the shock energy \cite{thompson04}.  
Depending on the angular distribution of the wind kinetic energy, 
the latter option requires that $\gtrsim10^{51}$\,ergs 
is extracted from the proto-magnetar on a $\lesssim1-2$\,s timescale
(see Ref \cite{bucciantini09,uzdensky_macfadyen}).

Proto-magnetars (and their GRBs) 
may also be formed by the accretion-induced collapse (AIC) of white 
dwarfs and (perhaps) by the merger of white dwarfs.  In this scenario 
there is no explosive nucleosynthesis, and perhaps little $^{56}$Ni yield \cite{M07a,MQT07}. 
A standard SN is not expected to accompany AIC due to 
the low total ejecta mass and modest quantity of
radioactive ejecta \cite{woosley_baron}.  However, 
Ref \cite{metzger09} show that the yield of
$^{56}$Ni in AIC is significantly enhanced in the case 
of rapid rotation due to winds from the accretion disk that
forms around the neutron star.  The radioactive decay 
of the ejecta produces an optical transient that is somewhat
dimmer and evolves much faster (on a timescale $\sim 1$ day) 
than a normal SN \cite{metzger09,darbha}, but which could in principle 
be detected with rapid and deep follow-up of a nearby
event. Ref \cite{dessart07} predict that magnetar 
formation via AIC ejects up to $\sim 0.1M_{\odot}$ in highly
neutron-rich material.  The heavy r-process elements 
produced in such ejecta contribute a comparable amount of
radioactive heating to $^{56}$Ni on timescales $\sim 1$ 
day \cite{metzger10} and thus may also contribute to
SN-like emission following the GRB.
 
Importantly, if proto-magnetar-driven GRBs 
can arise from AIC, then long-duration GRBs 
could trace both
young and old stellar populations.  This
is qualitatively different from the collapsar model.

\vspace*{-.5cm}
\subsection{Does Magnetar Birth Always Produce a GRB?}
\vspace*{-.2cm}

No.  Estimates suggest that $\sim10$\% of all
core-collapse SNe produce magnetars \cite{woods}.
Thus, the magnetar birth rate likely significantly exceeds 
the beaming-corrected GRB rate \cite{guetta}.  There are 
at least four possibilities for relieving this potential tension: 
(1) many magnetars are born slowly rotating and thus do not produce
GRBs, (2) many millisecond magnetars are born in Type-II progenitors
with extended hydrogen envelopes and their jets
are quenched \cite{zhang03}, (3) many millisecond magnetars simply do
not form jets able to escape the star (because of, e.g., instabilities), 
and finally (4) (a variant
on the first), because $\dot{E}$ (e.g., eq.~\ref{edot_ff}) is a very strong
function of both $B$ and $\Omega$, only the magnetars
born at the extremes of these distributions (e.g., $B\gtrsim10^{15.5}$\,G,
$P\lesssim2$\,ms) produce bright cosmological GRBs.  The rest give rise to 
less energetic events like XRFs or normal SNe.




\vspace*{-.3cm}
\begin{theacknowledgments}
We thank Eliot Quataert, Philip Chang, and Jonathan Arons for 
numerous stimulating discussions that motivated much of this proceedings.
T.A.T. is supported in part by an Alfred P.~Sloan Foundation Fellowship.
\end{theacknowledgments}




\bibliographystyle{aipproc}   




\end{document}